\documentclass[pra,aps,onecolumn,superscriptaddress,notitlepage]{revtex4-2}
\usepackage{amsmath,amssymb,bm}
\usepackage{mathtools}
\usepackage{graphicx}
\usepackage{epstopdf}
\usepackage{latexsym}
\usepackage[normalem]{ulem} 
\usepackage{appendix}
\usepackage{dsfont}
\usepackage{braket}
\usepackage{soul}
\usepackage{simpler-wick}
\usepackage{array}

\begin{document}

\newcommand{\Com}[1]{{\color{red}{#1}\normalcolor}} %Comment
\newcommand{\Rev }[1]{{\color{blue}{#1}\normalcolor}} % Revision
\newcommand{\mathcolorbox}[2]{\colorbox{#1}{$\displaystyle #2$}}

\newcommand{\ketbra}[2]{|#1\rangle\langle #2|}%ketbra
\newcommand{\normord}[1]{\mathopen{:}\,#1\,\mathopen{:}}

\author{Diego Barberena}
\email{diba2060@colorado.edu}
\affiliation{JILA, NIST and Department of Physics, University of Colorado, Boulder, Colorado 80309, USA}
\affiliation{Center for Theory of Quantum Matter, University of Colorado, Boulder, Colorado 80309, USA}
\author{Robert J. Lewis-Swan}
\affiliation{Homer L. Dodge Department of Physics and Astronomy, The University of Oklahoma, Norman, OK 73019, USA}
\affiliation{Center for Quantum Research and Technology, The University of Oklahoma, Norman, OK 73019, USA}
\author{Ana Maria Rey}
\affiliation{JILA, NIST and Department of Physics, University of Colorado, Boulder, Colorado 80309, USA}
\affiliation{Center for Theory of Quantum Matter, University of Colorado, Boulder, Colorado 80309, USA}
\author{James K. Thompson}
\affiliation{JILA, NIST and Department of Physics, University of Colorado, Boulder, Colorado 80309, USA}
\date{\today}

% \author{\firstname{Diego} \lastname{Barberena}\IsCorresp}
% \address{JILA, NIST, Department of Physics, University of Colorado,  Boulder, CO 80309, USA}
% \address{Center for Theory of Quantum Matter, University of Colorado, Boulder, CO 80309, USA}
% \email[D. Barberena]{diego.barberena@colorado.edu}
% \author{\firstname{Robert} \middlename{J.} \lastname{Lewis-Swan}}
% % \addressSameAs{1}{JILA, NIST, Department of Physics, University of Colorado,  Boulder, CO 80309, USA}
% % \addressSameAs{2}{Center for Theory of Quantum Matter, University of Colorado, Boulder, CO 80309, USA}
% \address{Homer L. Dodge Department of Physics and Astronomy, The University of Oklahoma, Norman, OK 73019, USA}
% \address{Center for Quantum Research and Technology, The University of Oklahoma, Norman, OK 73019, USA}
% \email[R. J. Lewis-Swan]{lewisswan@ou.edu}
% \author{\firstname{Ana Maria} \lastname{Rey}}
% \addressSameAs{1}{JILA, NIST, Department of Physics, University of Colorado,  Boulder, CO 80309, USA}
% \addressSameAs{2}{Center for Theory of Quantum Matter, University of Colorado, Boulder, CO 80309, USA}
% \email[A. M. Rey]{arey@jilau1.colorado.edu }
% \author{\firstname{James} \middlename{K.} \lastname{Thompson}}
% \addressSameAs{1}{JILA, NIST, Department of Physics, University of Colorado,  Boulder, CO 80309, USA}
% \email[J. K. Thompson]{jkt@jila.colorado.edu}

\title{Ultra narrow linewidth frequency reference via measurement and feedback}
%Possible titles
%Optical frequency stabilization using ultra-long-lived dipoles inside a QED cavity
%Noise and frequency response of coherently driven ultracold atoms inside an optical QED cavity
%Ultra narrow emission via cooperative resonance fluoresence 
%Ultra narrow linewidth light via cooperative resonance fluoresence

\date{\today}

\begin{abstract}
%We analyze the efficacy for frequency estimation and stabilization of an ensemble of atoms driven by coherent light inside an optical cavity.
The generation of very narrow linewidth light sources is of great importance in modern science. One such source is the superradiant laser, which relies on collectively interacting ultra long lived dipoles driven by incoherent light. Here we discuss a different way of generating spectrally pure light by coherently driving such dipoles inside an optical QED cavity. The light exiting the cavity carries information about the detuning between the driving light and the atomic transition, but is also affected by the noise originating from all the decoherence processes that act on the combined atom-cavity system. We calculate these effects to obtain fundamental limits for frequency estimation and stabilization across a range of values of input light intensities and atom-light interaction strengths, estimate these limits in state-of-the-art cavity experiments with alkaline-earth atoms and identify favorable operating conditions. We find that the achievable linewidths are comparable to those of the superradiant laser.
\end{abstract}


\maketitle  

\section{Introduction} 
Sources of coherent light with high frequency stability are crucial components of modern day technologies. In line with these requirements, the proposal for a continuous superradiant laser~\cite{Meiser2009,CHE09,Bohnet2012,Norcia2015b,norcia2016superradiance,Schaffer2020,Zhang2021,Kazakov,Cline2022} could provide an improved active frequency reference that is largely insensitive to cavity frequency noise, which limits modern day most stable lasers~\cite{Kessler2012}. Such a device comprises an electromagnetic mode of an optical cavity that is made to interact collectively with an ensemble of atoms that host a long lived two-level transition (ground and excited states). The atoms are incoherently driven by exciting the ground state to an auxiliary third level that decays quickly onto the excited state, and this is done in a regime where the cavity linewidth is much larger than the bare atomic linewidth. Above a critical value of the incoherent pump rate, the interplay between the incoherent drive and the atom-light interactions synchronize the atomic array forcing it to collectively emit coherent light with a linewidth set by the system cooperativity times the atomic linewidth. Although the superradiant laser can open great opportunities for quantum metrology applications, the need of a strong incoherent pump can significantly heat the atoms making the experimental implementation of continuous superradiance in optical transitions challenging. Up to date, superradiance has been realized in a pulsed way, in metrologically relevant transitions~\cite{Norcia2018Super,Laske2019}, or quasi-continuously, in transitions with broader linewidths~\cite{Norcia2016c}. 
%Up to date only pulsed superradiance has been realized~\cite{norcia2016superradiance,Norcia2018Super,Laske2019}. 
In this article, we analyze this same system in a complementary situation, where the long lived transition is excited directly with a laser. The light coming outside of the cavity can then be used to estimate the detuning between the driving light and the atomic transition, possibly allowing for frequency stabilization through a feedback scheme.

% and has been denoted cooperative resonance fluoresence conectar con el título. 
This system has been extensively analyzed in the past under the name of cooperative fluorescence, with studies focusing on its steady state properties~\cite{Carmichael_1980,Drummondo1980,Walls1980}, dynamics~\cite{Mielke:97,DRUMMOND1978}, bistability and hysteresis~\cite{BONIFACIO1976,Gripp1996}, correlation functions of the output light~\cite{Foster2000}, etc; and has seen a resurgence of interest from recent experiments~\cite{Ferioli2022} that probe the superradiant phase transition~\cite{Carmichael_1980,Walls1980} in free space. There has also been previous work in frequency stabilization at large input fields~\cite{Martin2011}, where the atomic ensemble is strongly saturated, or in situations where the atomic transition is affected by motion ~\cite{Westergaard2015,Tieri2015}. Here we give a comprehensive description of this system's ability to provide information on the frequency of the driving light across many values of input light intensity and effective atom-light interaction. Correctly assessing this requires two pieces: (1) the spectral response of the system to frequency fluctuations in the input light, and (2) the noise properties of the output light, which is affected by the presence of the atoms and is in general much noisier than that of a purely coherent light source.

Our results allow us to evaluate the advantages and disadvantages (for frequency stabilization) of operating the system in different parameter regimes. We determine the spectral purity of the emitted light after feedback stabilization and observe that it can be compatible with the one generated by a superradiant laser. Our analysis also leads us to the conclusion that artificially increasing the effective spontaneous emission by means of an extra depumping process may sometimes be advantageous in terms of frequency stability. 

%This paper is organized as follows: in section 2 we analyze the interaction of an ensemble of atoms with a cavity QED mode in the absence of single particle decoherence and in the presence of a constant laser drive [see Fig.~\ref{fig:Schematic}(a)]. We discuss the ideal case first because formulas are less cumbersome and the logic behind the steps can be explained more smoothly. In section 3 we expand the analysis to include the extra decoherence. This will lead to technical complications, but the spirit behind the calculations will be the same. The ideal case discussed in section 2 will also constitute a standard against which to compare later results.
\section{Ideal Model}
\begin{figure}
    \centering
    \includegraphics[width=0.9\textwidth]{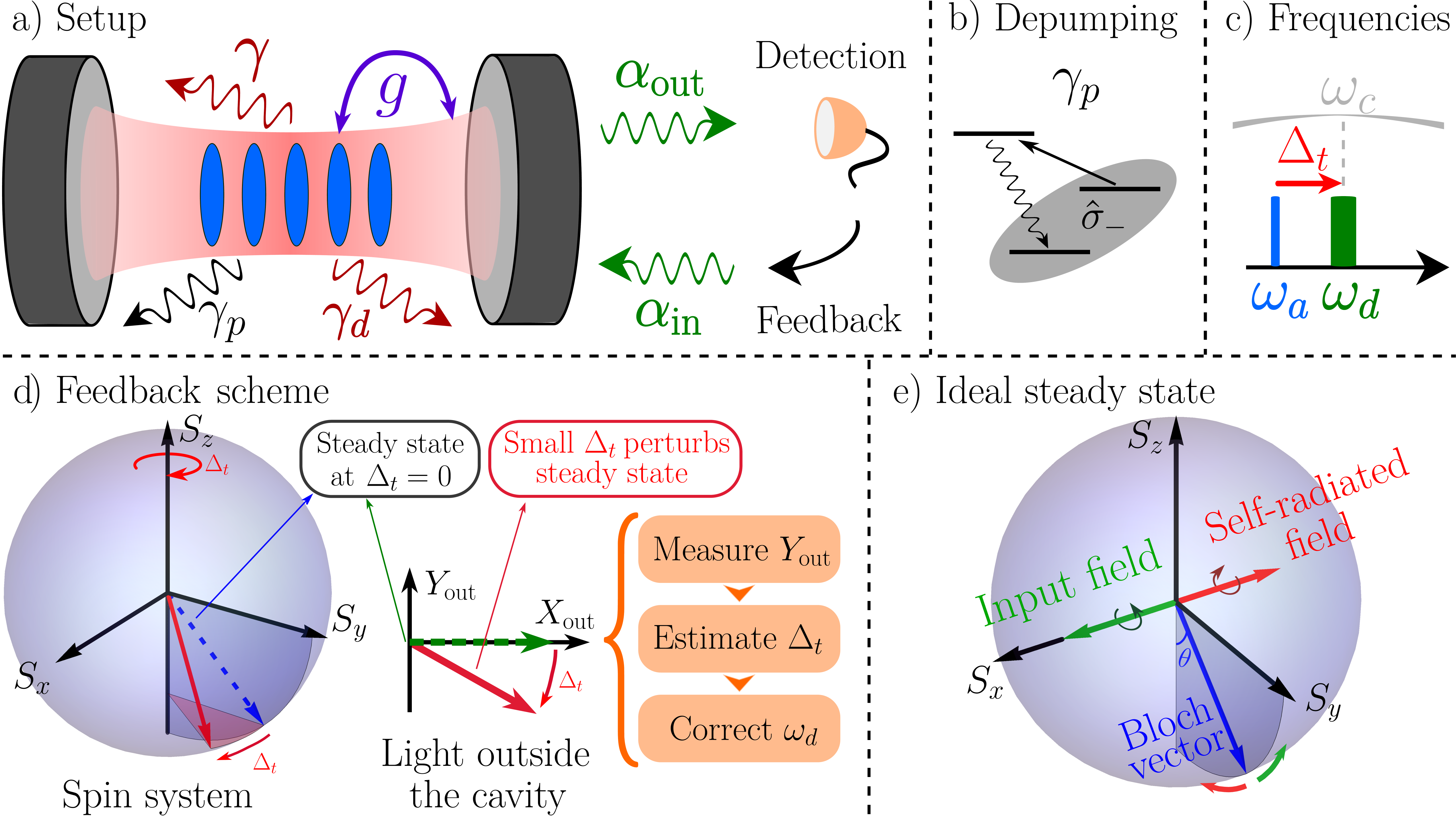}
    \caption{(a) Schematic of the system: the cavity is driven by coherent light ($\alpha_{\text{in}}$) and interacts with the atoms with strength $g$. The atoms are subject to dephasing ($\gamma_d$), spontaneous emission into free space ($\gamma$) and an additional depumping process [$\gamma_p$, see panel (b)]. The light coming out of the cavity ($\alpha_{\text{out}}$) is measured and this information is used to apply feedback on the input light source. (b) The depumping process can be engineered by means of a rapidly decaying auxiliary third level. (c) The cavity is frequency locked to the input light [$\omega_c=\omega_d(t)$], whose frequency fluctuates in time. They are detuned by $\Delta_t$ from the atomic transition. (d) Schematic of the feedback scheme: when $\Delta_t=0$ the system reaches a steady state. Small $\Delta_t$ perturbs this condition and this is manifested in atomic observables and the output light quadrature $Y_{\text{out}}$. (e) In the ideal system (without single particle decoherence), the steady state is characterized by a perfect cancellation between the torques induced by the input field and the self-radiated field of the atoms.}
    \label{fig:Schematic}
\end{figure}

In this section we will analyze the interaction of an ensemble of atoms with a cavity QED mode in the absence of single particle decoherence and in the presence of a constant laser drive [see Fig.~\ref{fig:Schematic}(a)]. We discuss the ideal case first because formulas are less cumbersome and the logic behind the steps can be explained more smoothly. The inclusion of the extra decoherence in section 3 will lead to technical complications, but the spirit behind the calculations will be the same. The ideal case will also constitute a standard with which to compare later results.

The analysis we will pursue here consists of four parts. First, we compute the non-equilibrium steady state of the system within the mean field approximation. Second, we calculate the linear response to two kinds of perturbations: (i) fluctuations in the input light frequency and (ii) quantum noise of the input light. Third, using the response of the output light to (i) we construct an estimator that allows us to infer the frequency of the driving light, while (ii) gives us the quantum noise in this estimator and provides a measure of its efficacy. We finalize by using (i) within a closed feedback loop to calculate the effective linewidth of the frequency stabilized input light source.

The system we are studying consists of an ensemble of $N$ two-level atoms with atomic transition $\omega_a$ collectively coupled to a cavity QED mode with resonance frequency $\omega_c$ and power decay linewidth $\kappa$ [see Fig.~\ref{fig:Schematic}(a)]. The Hamiltonian describing this system is (in units where $\hbar=1$):
\begin{equation}
    \hat{H}=\sum_{i=1}^N\frac{\omega_a}{2}(1+\sigma_i^z) +\omega_c\hat{a}^\dagger\hat{a}+g \sum_{i=1}^N(\hat{a}\hat{\sigma}_i^++\hat{a}^\dagger\hat{\sigma}_i^-),
\end{equation}
where $\hat{\sigma}_i^{x,y,z}$ are Pauli matrices describing the two-level system of atom $i$, $\hat{a}$ ($\hat{a}^\dagger$) is an annihilation (creation) operator describing the cavity mode and $2g$ is the single photon Rabi frequency. We further add a laser drive with instantaneous phase $\phi(t)$ [i.e. instantaneous frequency $\omega_d(t)=\dot{\phi}(t)$] and flux of $\alpha_{\text{in}}^2$ photons per second. To account for the properties of the light outside the cavity, we describe the evolution of the system in terms of Heisenberg-Langevin equations~\cite{Gardiner1985}, written in the lab frame as:
\begin{align}
    \begin{split}
        \partial_t\hat{a}&=-\bigg(i\omega_c+\frac{\kappa}{2}\bigg)\hat{a}-ig\hat{S}^-+\sqrt{\kappa}\big[\alpha_{\text{in}}e^{i\phi(t)}+\delta\hat{A}_{\text{in}}(t)\big]\\
        \partial_t\hat{S}^-&=-i\omega_a\hat{S}^-+2ig\hat{S}_z\hat{a}\\
        \partial_t\hat{S}_z&=ig(\hat{a}^{\dagger}\hat{S}^--\hat{a}\hat{S}^{+}),
    \end{split}
\end{align}
where $\hat{S}_{x,y,z}=\sum_{i=1}^N \hat{\sigma}_i^{x,y,z}/2$ are collective spin operators and $\delta\hat{A}_{\text{in}}(t)$ accounts for the quantum fluctuations of the input field. Formally, $\delta\hat{A}_{\text{in}}$ acts on the Hilbert space associated to the continuum of modes outside the cavity and is constructed of Heisenberg operators in the infinite past~\cite{Gardiner1985}. Given that the input state is coherent, it is entirely characterized by: $\langle\delta\hat{A}_{\text{in}}(t)\rangle=0$, and $\langle\delta\hat{A}_{\text{in}}(t)\delta\hat{A}_{\text{in}}^\dagger(t')\rangle=\delta(t-t')$, with all other second order expectations equal to 0. 
%These expectation values are evaluated with respect to the quantum state in the infinite past since we are working in the Heisenberg picture, where all time dependence is carried by the operators.
The output field, which describes the properties of the light outside of the cavity in the infinite future, can then be calculated using the input-output relation  $\hat{A}_{\text{out}}(t)=\alpha_{\text{in}}e^{i\phi(t)}+\delta\hat{A}_{\text{in}}(t)-\sqrt{\kappa}\hat{a}$.

In what follows we assume that the cavity is locked to the drive [$\omega_c=\omega_d(t)=\dot\phi(t)$] and close to resonance with the atomic transition [see Fig.~\ref{fig:Schematic}(c)]. In the rotating frame of the drive/cavity, the equations of motion take the simpler form:
\begin{align}\label{eqn:Heis-LangeIdeal}
    \begin{split}
        \partial_t\hat{a}&=-\frac{\kappa}{2}\hat{a}-ig\hat{S}^-+\sqrt{\kappa}\big[\alpha_{\text{in}}+\delta\hat{A}_{\text{in}}(t)\big]\\
        \partial_t\hat{S}^-&=i\Delta_t\hat{S}^-+2ig\hat{S}_z\hat{a}\\
        \partial_t\hat{S}_z&=ig(\hat{a}^{\dagger}\hat{S}^--\hat{a}\hat{S}^{+}),
    \end{split}
\end{align}
where $\Delta_t\equiv\dot{\phi}(t)-\omega_a=\omega_d(t)-\omega_a$ is the atom-drive detuning, which we assume to be small. Locking the drive to the atoms corresponds to having $\Delta_t=0$, but in a real laser $\Delta_t$  is a fluctuating quantity. We thus need to estimate it and correct it towards 0 [see Fig.~\ref{fig:Schematic}(d)].
% The first step is thus first to determine the non-equilibrium steady state of the system and then calculate its response to the small perturbation $\Delta_t$. We perform this first step within the mean field approximation, which is valid in the limit $N\to\infty$ due to the collective nature of the atom-light coupling.
\subsection{Mean field analysis}
The mean field equations of motion are obtained by replacing the operators ($\hat{a},\hat{S}^-,\hat{S}_z,\hat{A}_{\text{out}}$) by c-numbers ($\alpha,J,Z,\alpha_{\text{out}}$) and omitting $\delta\hat{A}_{\text{in}}(t)$ in Eq.~(\ref{eqn:Heis-LangeIdeal}). The only stable steady state solution at $\Delta_t=0$ occurs below a drive threshold, $\alpha_{\text{in}}\leq \alpha_{\text{in}}^c=gN/(2\sqrt{\kappa})$, and is characterized by:
\begin{equation}
    \alpha=0\hspace{0.75cm}J=-\frac{i N}{2}\sin\theta\hspace{0.75cm} Z=-\frac{N}{2}\cos\theta
\end{equation}
where $\sin\theta=\alpha_{\text{in}}/\alpha_{\text{in}}^c$. Physically, at nonzero inversion ($Z$) the atoms radiate a field into the cavity that precisely cancels the field generated by the input light. This leads to zero intracavity photons ($\alpha=0$) and a collective Bloch vector pointing at an angle ($\theta$) on the southern hemisphere, as shown in Fig.~\ref{fig:Schematic}(e). Using the input-output relation, we find that $\alpha_{\text{out}}=\alpha_{\text{in}}$, i.e. all the light is reflected. 

Above the threshold $\alpha_{\text{in}}^c$ there are no stable steady state solutions and mean field theory predicts persistent dynamical oscillations. The relaxation towards the steady state is caused by quantum-fluctuation-induced diffusion between classical trajectories, and leads, at very long times, to a very highly mixed steady state with exactly zero $Z$~\cite{Carmichael_1980}. This quantity is continuous but non-analytic at the threshold point, reminiscent of second order phase transitions, though notions of symmetry-breaking are related to other observables instead~\cite{Hannukainen2018,Link2019}.
\begin{figure}
    \centering
    \includegraphics[width=\textwidth]{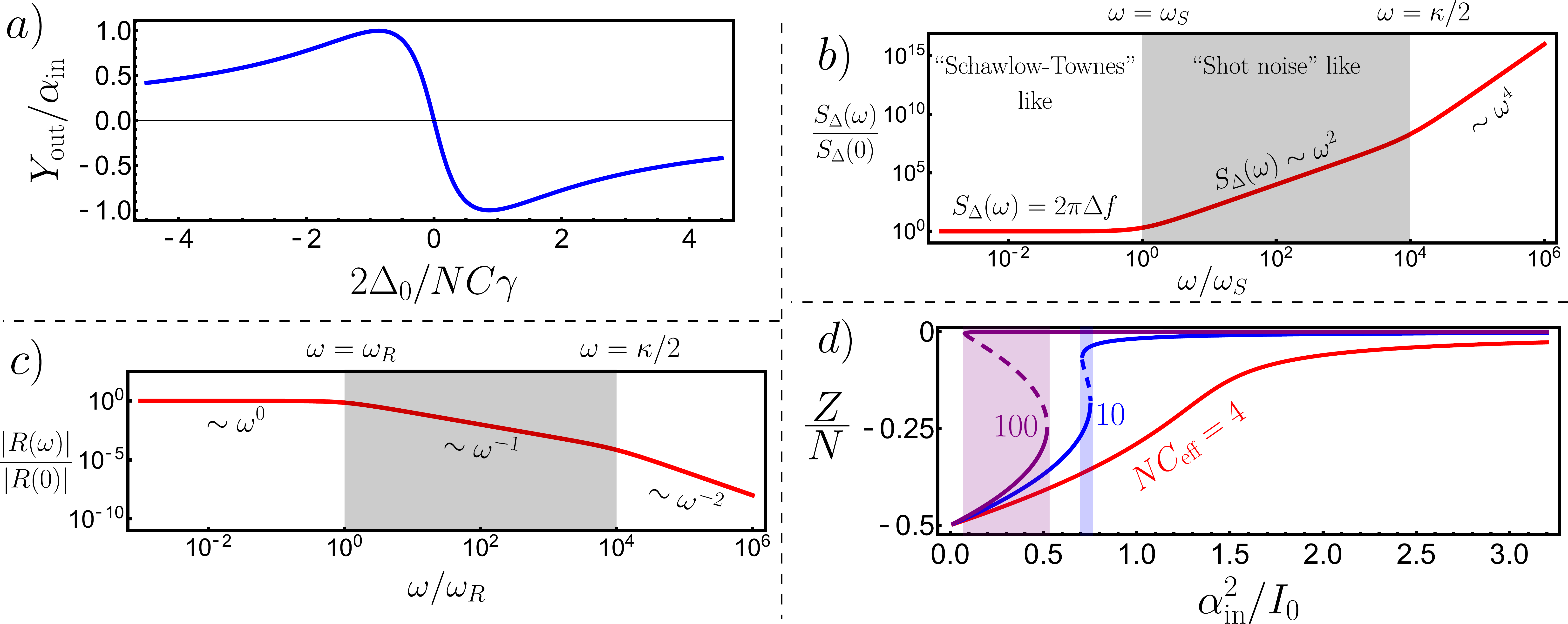}
    \caption{(a) Response of the system ($\hat{Y}$ quadrature of output light) to a static atom-drive detuning ($\Delta_0$) for fixed $\alpha_{\text{in}}=\alpha_{\text{in}}^c/\sqrt{2}$. (b) Representative spectral density of frequency fluctuations after feedback stabilization. (c) Representative response of the system to frequency fluctuations. (d) Steady state atomic inversion $Z/N$ as a function of input light intensity in the presence of single particle decoherence for three different values of effective cooperativity $NC_{\text{eff}}=4,10,100$ (red, blue and purple, respectively). Solid lines are stable solutions while dashed lines are unstable. Regions of bistability are shaded.}
    \label{fig:IdealSystem}
\end{figure}
\subsection{Fluctuations}
We now linearize Eq.~(\ref{eqn:Heis-LangeIdeal}) with respect to the mean field steady state solution. The equations are more easily written in terms of fluctuations of the intracavity field quadratures $\hat{X}=(\hat{a}+\hat{a}^\dagger)/2$, $\hat{Y}=(\hat{a}-\hat{a}^\dagger)/(2i)$, of input field quadratures $\delta\hat{X}_{\text{in}}=(\delta\hat{A}_{\text{in}}+\delta\hat{A}_{\text{in}}^\dagger)/2$, $\delta\hat{Y}_{\text{in}}=(\delta\hat{A}_{\text{in}}-\delta\hat{A}_{\text{in}}^\dagger)/(2i)$ and of spin fluctuations perpendicular to the mean field Bloch vector: $\hat{S}_x$ and $\hat{S}_\perp=\hat{S}_y\cos\theta+\hat{S}_z\sin\theta$. This results in the following sets of decoupled linear equations:
% \begin{align}
%     \begin{split}
%         \partial_t\delta\hat{a}&=-\frac{\kappa}{2}\delta\hat{a}-ig \delta\hat{S}+\sqrt{\kappa}\delta\hat{A}_{\text{in}}(t)\\
%         \partial_t\delta\hat{S}&=(2igZ)\delta\hat{a}-\nu_t\frac{N}{2}\frac{\alpha_{\text{in}}}{\alpha_{\text{in}}^c}\\
%         \partial_t\delta\hat{S}_z&=\frac{gN}{2}\bigg(\frac{\alpha_{\text{in}}}{\alpha_{\text{in}}^c}\bigg)(\delta\hat{a}^\dagger+\delta\hat{a})
%     \end{split}
% \end{align}
\begin{equation}
\partial_t \begin{pmatrix} \hat{S}_x\\ \hat{Y} \end{pmatrix}=\begin{pmatrix} 0& gN\cos\theta\\ -g & -\frac{\kappa}{2}\end{pmatrix}\begin{pmatrix} \hat{S}_x\\ \hat{Y}\end{pmatrix}+\begin{pmatrix}\frac{N}{2}\Delta_t \sin\theta\\ \sqrt{\kappa}\delta\hat{Y}_{\text{in}}\end{pmatrix},\hspace{0.75cm} \partial_t \begin{pmatrix} \hat{S}_\perp\\ \hat{X} \end{pmatrix}=\begin{pmatrix} 0& gN\\ -g\cos\theta & -\frac{\kappa}{2}\end{pmatrix}\begin{pmatrix} \hat{S}_\perp\\ \hat{X}\end{pmatrix}+\begin{pmatrix}0\\ \sqrt{\kappa}\delta\hat{X}_{\text{in}}\end{pmatrix}.
\end{equation}
The direction parallel to the mean field Bloch vector, $\hat{S}_{||}=\hat{S}_z\cos\theta-\hat{S}_y\sin\theta$, is only sensitive to fluctuations at second order due to the curvature of the Bloch sphere [$\delta\hat{S}_{||}=1/2-(\hat{S}_x^2+\hat{S}_{\perp}^2)/N$]. From the linearized equations and the input-output relation we can obtain expressions for the Fourier modes of the output field, defined as $\hat{A}_{\text{out}}(\omega)=\int e^{i\omega t}\hat{A}_{\text{out}}(t)\,dt$, with similar definitions for $\Delta(\omega),\,\delta\hat{A}_{\text{in}}(\omega),\,\hat{a}(\omega)$ and $\hat{S}_{x,y,z}(\omega)$. Omitting transient effects~\footnote{The expression for the output signal $\hat{Y}_{\text{out}}(\omega)$ is seemingly missing an independent contribution coming from atomic projection noise, but this is an incorrect assertion. Atomic projection noise in the steady state is entirely determined by dynamical noise ($\delta\hat{A}_{\text{in}}$), a fact that is further established by computing the steady state value of $\hat{S}_x(\omega)$ and noting that the correct variance of $\hat{S}_x$ (in agreement with~\cite{Carmichael_1980,Drummondo1980,Walls1980}) is recovered using correlation functions of $\delta\hat{A}_{\text{in}}$.}:
\begin{align}\begin{split}\label{eqn:IdealFluctuations}
    % \hat{X}_{\text{out}}(\omega)&=2\pi \alpha_{\text{in}}\delta(\omega)+\frac{\big(i\omega+\lambda_+\big)\big(i\omega+\lambda_-\big)}{\big(i\omega-\lambda_+\big)\big(i\omega-\lambda_-\big)}\delta\hat{X}_{\text{in}}(\omega)\\[3pt]
    \hat{Y}_{\text{out}}(\omega)&=-\frac{\big(gN\sin\theta\sqrt{\kappa}/2\big)}{\big(i\omega-\lambda_+\big)\big(i\omega-\lambda_-\big)}\Delta(\omega)+\frac{\big(i\omega+\lambda_+\big)\big(i\omega+\lambda_-\big)}{\big(i\omega-\lambda_+\big)\big(i\omega-\lambda_-\big)}\delta\hat{Y}_{\text{in}}(\omega),
%   \hat{S}_x(\omega)&=-\frac{\big(i\omega-\kappa/2\big)\big(N\sin\theta/2\big)}{\big(i\omega-\lambda_+\big)\big(i\omega-\lambda_-\big)}\Delta(\omega)+\frac{gN\sqrt{\kappa}\cos\theta }{\big(i\omega-\lambda_+\big)\big(i\omega-\lambda_-\big)}\delta\hat{Y}_{\text{in}}(\omega),
\end{split}\end{align}
where $\lambda_{\pm}=(\kappa/4)\pm\sqrt{(\kappa/4)^2-g^2N\cos\theta}$. The output light quadrature $\hat{Y}_{\text{out}}$ has two contributions: one proportional to $\Delta(\omega)=\int e^{i\omega t}\Delta _t\,dt$ (signal) and another one proportional to $\delta\hat{Y}_{\text{in}}(\omega)$ (noise).

The real parts of $\lambda_\pm$ are the decay constants of the system. The smallest one determines how long it takes the system to relax to the steady state. When $\alpha_{\text{in}}\to\alpha_{\text{in}}^c$ ($\theta\to\pi/2$), this timescale increases without bounds, rendering the steady state condition harder to achieve. In the limit that $g\sqrt{N}\ll \kappa$, relevant for narrow linewidth transitions, these constants become $\lambda_+\approx \kappa/2$ and $\lambda_-\approx 2g^2N\cos\theta/\kappa$ (thus $\lambda_+\gg \lambda_-$). The combination $g^2/\kappa$ appears often and is typically rewritten as $4g^2/\kappa\equiv C\gamma$, where  $\gamma$ is the natural linewidth of the transition and $C=4g^2/(\kappa\gamma)$ is the cooperativity, which depends only on cavity geometry.

Equations~(\ref{eqn:IdealFluctuations}) relate the fluctuations of quantum observables in the steady state to the fluctuations of the input field ($\delta\hat{X}_{\text{in}},\,\delta\hat{Y}_{\text{in}}$). For example, the variance of $\hat{Y}_{\text{out}}$ defines $S_{Y_{\text{out}}}(\omega)$, the noise spectral density of the output light, according to:
\begin{equation}
    \langle \delta\hat{Y}_{\text{out}}(\omega)\delta\hat{Y}_{\text{out}}(\omega')\rangle\equiv \frac{2\pi S_{Y_{\text{out}}}(\omega)\,\delta(\omega+\omega')}{4}.
\end{equation}
Using $\langle\delta\hat{Y}_{\text{in}}(\omega)\delta\hat{Y}_{\text{in}}(\omega')\rangle=2\pi\delta(\omega+\omega')/4$ and $\langle\delta\hat{Y}_{\text{in}}(\omega)\rangle=0$, it follows that $S_{Y_{\text{out}}}(\omega)=1$, which is the value associated to shot noise. A similar result holds for $S_{X_{\text{out}}}(\omega)$, which indicates that the output light is in a coherent state~\cite{Somech2022}. 

% We can also calculate atomic observables to recover known results such as the steady state variance of $\hat{S}_x$ at $\Delta_t=0$~\cite{Carmichael_1980,Barberena2019}
% \begin{equation}
%     \langle\hat{S}_x^2\rangle=\int \frac{d\omega}{2\pi} \frac{d\omega'}{2\pi}\braket{\hat{S}_x(\omega)\hat{S}_x(\omega')}e^{-i(\omega+\omega')t}=\int\frac{d\omega}{2\pi}\frac{g^2N^2\kappa(\cos\theta)^2/4}{\big(\omega^2+\lambda_+^2\big)\big(\omega^2+\lambda_-^2\big)}=\frac{N\cos\theta}{4},
% \end{equation}
% extended here to any value of $g\sqrt{N}/\kappa$ and depicted in Fig.~\ref{fig:IdealSystem}(b). Close to threshold, the linearization analysis breaks down and curvature effects modify all the previous results. Hence, $\theta$ cannot get arbitrarily close to $\pi/2$.
\subsection{Frequency estimation}
A glance at Eq.~(\ref{eqn:IdealFluctuations}) indicates that the steady state output light quadrature $\hat{Y}_{\text{out}}$ carries information about the drive detuning $\Delta_t$. This suggests that this system can be used as a passive frequency reference, similar in spirit to Ramsey spectroscopy, where an observable carrying information about $\Delta_t$ is measured after an operation time $T$. In the presence of a static detuning $\Delta_t=\Delta_0$ [$\Delta(\omega)=2\pi\Delta_0\delta(\omega)$], the most straightforward observable that serves as an estimator of $\Delta_0$ is constructed from the average of $\hat{Y}_{\text{out}}$ over time. Within the linear regime:
\begin{equation}
    \hat{Y}_{\text{est}}\equiv\frac{1}{T}\int_0^T\hat{Y}_{\text{out}}(t)\,dt\longrightarrow \langle\hat{Y}_{\text{est}}\rangle=\frac{\Delta_0}{2g\cot\theta/\sqrt{\kappa}}.
\end{equation}
This linear relation is valid for $\Delta_0\lesssim NC\gamma$ [condition obtained by solving the mean field equations for finite static $\Delta_t=\Delta_0$, see Fig.~\ref{fig:IdealSystem}(a)], beyond which a typical dispersive response is observed. A real measurement of $\hat{Y}_{\text{est}}$ over a finite time is subject to quantum noise coming from the fluctuations of the output light, which are related to $\delta\hat{Y}_{\text{in}}(\omega)$ in Eq.~(\ref{eqn:IdealFluctuations}). The estimator variance after an integration time $T$ is calculated using Eq.~(\ref{eqn:IdealFluctuations}):
\begin{align}
    \begin{split}
        \langle\delta\hat{Y}_{\text{est}}^2\rangle&\equiv \big\langle (\hat{Y}_{\text{est}}-\langle\hat{Y}_{\text{est}}\rangle)^2\big\rangle=\frac{S_{\text{Y}_{\text{out}}}(0)}{4T}=\frac{1}{4T},
    %     &=\int_0^Tdt\int_0^T dt'\int \frac{d\omega}{2\pi T}\frac{d\omega'}{2\pi T}e^{-i(\omega t+\omega' t')}\\
    % &\frac{\big(i\omega+\kappa/2\big)\big(i\omega+2g^2N\cos\theta/\kappa\big)}{\big(-i\omega+\kappa/2\big)\big(-i\omega+2g^2N\cos\theta/\kappa\big)}\frac{\big(i\omega'+\kappa/2\big)\big(i\omega'+2g^2N\cos\theta/\kappa\big)}{\big(-i\omega'+\kappa/2\big)\big(-i\omega'+2g^2N\cos\theta/\kappa\big)}\langle\delta\hat{Y}_{\text{in}}(\omega)\delta\hat{Y}_{\text{in}}(\omega')\rangle\\
    % &=\frac{1}{4T}
    \end{split}
\end{align}
and from this we can compute the sensitivity:
\begin{equation}
    \delta\Delta_0^2=\frac{\langle\delta\hat{Y}_{\text{est}}^2\rangle}{(\partial_{\Delta_0}\langle\hat{Y}_{\text{est}}\rangle)^2}=\frac{g^2(\cot\theta)^2}{\kappa T}=\frac{C\gamma (\cot\theta)^2}{4 T}.
\end{equation}
As an example, consider the $^1S_0\to ^3\hspace{-0.1cm}P_0$ transition in $^{87}$Sr ($\lambda_a=698$ nm) after one second integration time and operating at $\theta=45^\circ$. Using the cavity parameters in~\cite{norcia2018cavity,Norcia2016c}, this corresponds to a relative frequency resolution of $\delta\Delta_0/\omega_a \approx 10^{-17}$. Experimentally, such a measurement of $\hat{Y}_{\text{out}}$ is achieved by means of a homodyne detection setup where the input light field acts as the local oscillator.
\subsection{Stabilization via feedback}\label{sec:feed}
The light coming outside of the cavity can be used within a closed feedback loop to provide automatic frequency stabilization. As shown in the previous section, we can estimate the detuning between the laser drive and the atomic transition, and we can then use this information to correct the frequency of the laser drive. However, the output light is a fluctuating quantity and hence the feedback process introduces noise into the corrected frequency. The objective in this section is to estimate the size of this noise, following the discussion in Ref.~\cite{Riehle2004}.

We begin with the expression for the output quadrature $\hat{Y}_{\text{out}}(\omega)$ given in Eq.~(\ref{eqn:IdealFluctuations}):
\begin{equation}\label{eqn:YtoR}
    \hat{Y}_{\text{out}}(\omega)=\underbrace{-\frac{\big(gN\sin\theta\sqrt{\kappa}/2\big)}{\big(i\omega-\lambda_+\big)\big(i\omega-\lambda_-\big)}}_{R(\omega)}\hat{\Delta}(\omega)+\underbrace{\frac{\big(i\omega+\lambda_+\big)\big(i\omega+\lambda_-\big)}{\big(i\omega-\lambda_+\big)\big(i\omega-\lambda_-\big)}\delta\hat{Y}_{\text{in}}(\omega)}_{\hat{N}(\omega)},
\end{equation}
where $R(\omega)$ is the response of the signal $\hat{Y}_{\text{out}}(\omega)$ to the perturbation $\hat{\Delta}(\omega)$, and $\hat{N}(\omega)$ describes the quantum fluctuations of $\hat{Y}_{\text{out}}(\omega)$. Note also that we are now writing $\hat{\Delta}(\omega)$ as an operator because we are considering the steady state of the feedback process, where the input $\hat{\Delta}$ already includes the effects of quantum noise. 

We assume the feedback signal is obtained from the output light using a linear filter $\beta(\omega)$, which gives us a second equation $\hat{\Delta}_{\text{feed}}(\omega)=\beta(\omega)\hat{Y}_{\text{out}}(\omega)$. Finally, we have a third relation connecting the bare detuning in the absence of feedback [$\Delta_{\text{bare}}(\omega)$] to the corrected detuning: $\hat{\Delta}(\omega)=\Delta_{\text{bare}}(\omega)-\hat{\Delta}_{\text{feed}}(\omega)$. From these three equations we can obtain a direct relation between $\hat{\Delta}(\omega)$ and $\Delta_{\text{bare}}(\omega)$:
\begin{equation}
    \hat{\Delta}(\omega)=\frac{1}{1+R(\omega)\beta(\omega)}\Delta_{\text{bare}}(\omega)-\frac{\beta(\omega)}{1+R(\omega)\beta(\omega)}\hat{N}(\omega)\approx \frac{1}{R(\omega)\beta(\omega)}\Delta_{\text{bare}}(\omega)-\frac{\hat{N}(\omega)}{R(\omega)},
\end{equation}
where in the last equality we have assumed $|\beta(\omega)R(\omega)|\gg 1$ for the frequencies of interest (i.e. below the unity gain frequency of the feedback loop). If the original laser drive is already noisy, then $\Delta_{\text{bare}}(\omega)$ becomes stochastic. However, the feedback process suppresses this noise by the large factor $R(\omega)\beta(\omega)$, while at the same time introducing the extra noise $\delta\hat{\Delta}(\omega)=-\hat{N}(\omega)/R(\omega)$:
\begin{equation}
    \delta\hat{\Delta}(\omega)\approx\frac{\big(i\omega+\lambda_+\big)\big(i\omega+\lambda_-\big)}{gN\sqrt{\kappa}\sin\theta/2}\hat{Y}_{\text{in}}(\omega)
\end{equation}
with second order average:
\begin{equation}
    \langle\delta\hat{\Delta}(\omega)\delta\hat{\Delta}(\omega')\rangle=2\pi\delta(\omega+\omega')\Bigg[\frac{ \big(\omega^2+\lambda_+^2\big)\big(\omega^2+\lambda_-^2\big)}{g^2N^2\kappa\sin\theta^2}\Bigg].
\end{equation}
The coefficient in front of $2\pi\delta(\omega+\omega')$ defines the spectral density of frequency fluctuations $S_{\Delta}(\omega)$~\cite{Riehle2004,Martin2011}. In the limit $\kappa\gg g\sqrt{N}$, this will be of the form: 
% \begin{equation}\label{eqn:idealLW}
%     S_{\Delta}(\omega)=\underbrace{\frac{C\gamma \cot\theta^2}{4}}_{2\pi\Delta \hspace{-0.05cm}f} \Bigg[1+\bigg(\frac{\omega}{\kappa/2}\bigg)^2\Bigg]\Bigg[1+\bigg(\frac{\omega}{NC\gamma\cos\theta /2}\bigg)^2\Bigg] 
% \end{equation}
\begin{equation}\label{eqn:idealLW}
    S_{\Delta}(\omega)=2\pi\Delta \hspace{-0.05cm}f \Bigg[1+\bigg(\frac{\omega}{\kappa/2}\bigg)^2\Bigg]\Bigg[1+\Bigg(\frac{\omega}{\omega_S}\Bigg)^2\Bigg], 
\end{equation}
with $\omega_S\ll \kappa/2$ [see Fig.~\ref{fig:IdealSystem}(b)]. The value of $S_{\Delta}(\omega)$ at $\omega=0$ determines the effective linewidth $\Delta \hspace{-0.05cm}f$. In this ideal scenario, $\Delta \hspace{-0.05cm}f=C\gamma \cot\theta^2/4$, which is the same scaling as the one that appears in studies of the superradiant laser~\cite{Meiser2009}. As $\theta\to\pi/2$ ($\alpha_{\text{in}}\to\alpha_{\text{in}}^c$), $\Delta \hspace{-0.05cm}f$ approaches 0, and thus it is ultimately limited by finite size effects. The corner frequency $\omega_S=NC\gamma \cos\theta/2$ marks the point at which $S_{\Delta}(\omega)$ switches from constant to $\sim\omega^2$ behaviour, which in a real laser are characteristic of the Schawlow-Townes noise floor and of shot noise (due to a finite laser output power) respectively.

The construction of the feedback loop [i.e. the specific choice of $\beta(\omega)$] also requires information on the response function $|R(\omega)|$ to guarantee large low-frequency gain [$R(0)\beta(0)$] and keep the loop stable. In this case, $|R(\omega)|$ behaves as [see Eq.~\ref{eqn:YtoR}]
\begin{equation}\label{eqn:Response}
    |R(\omega)|\propto\frac{1}{\big[1+(2\omega/\kappa)^2\big]^{1/2}\big[1+(\omega/\omega_R)^2\big]^{1/2}}
\end{equation}
and is schematically depicted in Fig.~\ref{fig:IdealSystem}(c). In this ideal case, $\omega_R$ is also $NC\gamma\cos\theta/2$, though this is not generic. Furthermore, a realistic calculation of $\Delta \hspace{-0.05cm}f$, $\omega_S$ and $\omega_R$ requires that single particle sources of decoherence be included in the analysis.
%suggesting that both modes of operation are subject to the same limitations
\section{Non ideal model}
In any realistic system there is dephasing and incoherent decay, and these processes will modify both the steady state of the system and its linear response to perturbations. This is encoded in the Heisenberg-Langevin equations that replace Eq.~(\ref{eqn:Heis-LangeIdeal}):
\begin{align}\label{eqn:Heis-LangeNonIdeal}
    \begin{split}
        \partial_t\hat{a}&=-\frac{\kappa}{2}\hat{a}-ig\hat{S}^-+\sqrt{\kappa}\big[\alpha_{\text{in}}+\delta\hat{A}_{\text{in}}(t)\big]\\
        \partial_t\hat{S}^-&=i\Delta_t\hat{S}^-+2ig\hat{S}_z\hat{a}-\bigg(\frac{\gamma+\gamma_d+\gamma_p}{2}\bigg)\hat{S}^-+\sqrt{N\gamma}\hat{F}_{\gamma}^-(t)+\sqrt{N\gamma_{p}}\hat{F}_{\gamma_p}^-(t)+\sqrt{N\gamma _d}\hat{F}_{\gamma_d}^-(t)\\
        \partial_t\hat{S}_z&=ig(\hat{a}^{\dagger}\hat{S}^--\hat{a}\hat{S}^{\dagger})-(\gamma+\gamma_p)\bigg(\hat{S}_z+\frac{N}{2}\bigg)+\sqrt{N\gamma }\hat{F}_{\gamma}^z(t)+\sqrt{N\gamma_p}\hat{F}_{\gamma_p}^z(t),
    \end{split}
\end{align}
where $\gamma_d$ is the dephasing rate, $\gamma$ is the natural spontaneous emission rate of the transition, $\gamma_p$ is the rate of the depumping process depicted in Fig.~\ref{fig:Schematic}(b), $\hat{F}_{\gamma}^-$ and $\hat{F}_{\gamma}^z$ are noise operators associated to spontaneous emission (notice $\hat{F}_{\gamma}^z$ is hermitian), and $\hat{F}_{\gamma_d}^-$ is a noise operator associated to dephasing. We have also included an additional depumping process, $\gamma_p$, with associated noise operators $\hat{F}_{\gamma_p}^-$ and $\hat{F}_{\gamma_p}^z$, and which can be engineered via coherent driving onto a rapidly decaying auxiliary third level [see Fig.~\ref{fig:Schematic}(b)]. In principle the noise operators are defined in terms of the degrees of freedom which cause the associated decay processes, but they are in general difficult to access, unlike $\hat{A}_{\text{out}}(t)$. In practice, we are interested mostly in their effect on the system, which is encoded in their lowest order correlators. The only nonzero ones (up to hermitian conjugation) are~\cite{Tieri2017,Meystre2007}:
\begin{equation}
    \begin{matrix*}[l]
    \langle \hat{F}_{\gamma}^-(t)\hat{F}_\gamma^+(t')\rangle&=\langle \hat{F}^-_{\gamma_p}(t)\hat{F}_{\gamma_p}^+(t')\rangle&=\delta(t-t')\\[2pt]
    \langle \hat{F}_\gamma^-(t)\hat{F}_{\gamma}^z(t')\rangle&=\langle \hat{F}_{\gamma_p}^-(t)\hat{F}_{\gamma_p}^z(t')\rangle&=\frac{\langle\hat{S}^-\rangle}{N}\delta(t-t')\\[2pt]
    \langle \hat{F}_{\gamma}^z(t)\hat{F}_{\gamma}^z(t')\rangle&=\langle \hat{F}_{\gamma_p}^z(t)\hat{F}_{\gamma_p}^z(t')\rangle&=\bigg(\frac{\langle\hat{S}_z\rangle}{N}+\frac{1}{2}\bigg)\delta(t-t')
    \end{matrix*}\hspace{1cm}
    \begin{matrix*}[l]
    \langle \hat{F}^-_{\gamma_d}(t)\hat{F}_{\gamma_d}^+(t')\rangle&=\bigg(\frac{1}{2}-\frac{\langle\hat{S}_z\rangle}{N}\bigg)\delta(t-t')\\[4pt]
    \langle \hat{F}^+_{\gamma_d}(t)\hat{F}_{\gamma_d}^-(t')\rangle&=\bigg(\frac{1}{2}+\frac{\langle\hat{S}_z\rangle}{N}\bigg)\delta(t-t').
    \end{matrix*}
\end{equation}
Similarly to the ideal case, this system hosts resonant cooperative fluorescence~\cite{BONIFACIO1976}. As in the previous section, we first investigate the system within the mean field approximation, then compute the fluctuations of the output light with respect to the steady state and analyze how these incoherent processes modify our results for frequency estimation and feedback.
\subsection{Mean field analysis}
Once again, we replace ($\hat{a},\hat{S}^-,\hat{S}_z,\hat{A}_{\text{out}}$) by c-numbers ($\alpha,J,Z,\alpha_{\text{out}}$) and omit all noise operators. The resulting equations are nonlinear but their steady state solution can be parameterized in terms of $z=Z/N$ as:
\begin{equation}\label{eqn:NonIdealSS}
    \alpha=\frac{2\alpha_{\text{in}}}{\sqrt{\kappa}\big(1-2NC_{\text{eff}}z\big)}\hspace{0.5cm}\frac{J}{N}=i\sqrt{\frac{\gamma+\gamma_p}{\Gamma}}\Bigg(\frac{\alpha_\text{in}}{\sqrt{16 I_0}}\Bigg)\Bigg(\frac{2NC_{\text{eff}}z}{1-2NC_{\text{eff}}z}\Bigg)\hspace{0.5cm} \alpha_{\text{out}}=\alpha_{\text{in}}\Bigg(\frac{2zNC_{\text{eff}}+1}{2zNC_{\text{eff}}-1}\Bigg),
\end{equation}
where $C_{\text{eff}}\equiv4g^2/(\Gamma\kappa)=C \gamma/\Gamma$ is the effective cavity cooperativity, $\Gamma=\gamma+\gamma_d+\gamma_p$ is the total dephasing rate, $I_0=N^2C_{\text{eff}}(\gamma+\gamma_p)/16$ and $z$ satisfies:
\begin{equation}\label{eqn:NonIdealZ}
    \frac{\alpha_{\text{in}}^2}{8I_0}=-\frac{(z+1/2)}{z}\bigg(z-\frac{1}{2NC_{\text{eff}}}\bigg)^2.
\end{equation}
The natural scale for $\alpha_{\text{in}}^2$ is set by $I_0$ and is connected to the ideal threshold field ($\alpha_{\text{in}}^c$) by $I_0=(\alpha_{\text{in}}^c)^2(\gamma+\gamma_p)/\Gamma$. The fate of the transition present in the ideal model depends on the value of $NC_{\text{eff}}$. When $NC_{\text{eff}}<8$, there is only one real solution to Eq.~(\ref{eqn:NonIdealZ}) and the mean field observables are smooth functions of the input field strength $\alpha_{\text{in}}$ [see Fig.~\ref{fig:IdealSystem}(d) for $NC_{\text{eff}}=4$]. Notably, there is one ``dark point" where $\alpha_{\text{out}}=0$, i.e. no light comes out of the cavity [see Eq.~(\ref{eqn:NonIdealSS}) with $z=-1/(2NC_{\text{eff}})$]. In the context of cooperative resonance fluorescence, as discussed before in Ref.~\cite{Carmichael_1980,BONIFACIO1976,Walls1980}, the smooth behavior observed in this regime is a manifestation of the disappearance of the phase transition observed in the absence of decoherence.

When $NC_{\text{eff}}>8$, there are input field amplitudes $\alpha_{\text{in}}$ for which there are three real solutions for $z$. Two of them correspond to stable steady states and one is in an unstable branch [see Fig.~\ref{fig:IdealSystem}(d) for $NC_{\text{eff}}=10,100$]~\cite{BONIFACIO1976}. This leads to bistability when:
\begin{equation}
    \frac{1}{16}\Bigg(1-\sqrt{1-\frac{8}{NC_{\text{eff}}}}\Bigg)\Bigg(3+\sqrt{1-\frac{8}{NC_{\text{eff}}}}\Bigg)^3<\frac{2\alpha_{\text{in}}^2}{I_0}<\frac{1}{16}\Bigg(1+\sqrt{1-\frac{8}{NC_{\text{eff}}}}\Bigg)\Bigg(3-\sqrt{1-\frac{8}{NC_{\text{eff}}}}\Bigg)^3
\end{equation}
In the limit $NC_{\text{eff}}\gg 1$, the lower and upper bounds become $16/(NC_{\text{eff}})$ and $1$~\cite{Tucker2020}, respectively. This bistable behaviour is accompanied by large discontinuous jumps between stable solutions and thus the presence of hysteresis~\cite{BONIFACIO1976} instead of the smooth second order transition observed in the ideal case~\cite{Carmichael_1980}. It is preferable to work in non-bistable regimes to avoid the large fluctuations associated with these switches.

% \begin{figure}
%     \centering
%     \includegraphics[width=\textwidth]{Figures/Fig3_rescaled.pdf}
%     \caption{(a) Steady state atomic inversion $Z/N$ as a function of input light intensity in the presence of single particle decoherence for three different values of effective cooperativity $NC_{\text{eff}}=4,10,100$ (red, blue and purple, respectively). Solid lines are stable solutions while dashed lines are unstable. Regions of bistability are shaded. (b) Output light amplitude.}
%     \label{fig:NonIdealMeanField}
% \end{figure}
\subsection{Fluctuations}
We linearize Eq.(\ref{eqn:Heis-LangeNonIdeal}) about any of the mean field steady state solutions and look at $\hat{Y}$, $\hat{Y}_{\text{out}}$ and $\hat{S}_x$, which carry information about $\Delta_t$. They satisfy:
\begin{equation}
\partial_t \begin{pmatrix} \hat{S}_x\\ \hat{Y} \end{pmatrix}=\begin{pmatrix} -\frac{\Gamma}{2}& -2gNz\\ -g & -\frac{\kappa}{2}\end{pmatrix}\begin{pmatrix} \hat{S}_x\\ \hat{Y}\end{pmatrix}+\begin{pmatrix}(iJ)\Delta_t+\sqrt{N\gamma}\hat{X}_{\gamma}(t)+\sqrt{N\gamma_p}\hat{X}_{\gamma_p}(t)+\sqrt{N\gamma_d}\hat{X}_{\gamma_d}(t)\\ \sqrt{\kappa}\delta\hat{Y}_{\text{in}}\end{pmatrix},
\end{equation}
where $\hat{F}_{\gamma}^x=(\hat{F}_{\gamma}^-+\hat{F}_{\gamma}^+)/2$, $\hat{F}_{\gamma_p}^x=(\hat{F}_{\gamma_p}^-+\hat{F}_{\gamma_p}^+)/2$ and $\hat{F}_{\gamma_d}^x=(\hat{F}_{\gamma_d}^-+\hat{F}_{\gamma_d}^+)/2$ are the corresponding quadrature noise operators. We can then calculate the behaviour of $\hat{Y}_{\text{out}}$ through $\hat{Y}_{\text{out}}=\delta\hat{Y}_{\text{in}}-\sqrt{\kappa}\hat{Y}$. In frequency space, omitting transient behaviour:
\begin{align}\label{eqn:NonIdealYout}
    \begin{split}
        \hat{Y}_{\text{out}}(\omega)&=-\frac{2g\sqrt{\kappa}(iJ)}{\big(i\omega-l_+\big)\big(i\omega-l_-\big)}\Delta(\omega)+\frac{\big(i\omega+m_+\big)\big(i\omega+m_-)}{\big(i\omega-l_+\big)\big(i\omega-l_-\big)}\delta\hat{Y}_{\text{in}}(\omega)\\[3pt]
        &\hspace{2cm}+\frac{\big(\kappa/2\big)\sqrt{NC_{\text{eff}}\Gamma}}{\big(i\omega-l_+\big)\big(i\omega-l_-\big)}\big[\sqrt{\gamma}\hat{F}_{\gamma}^x(\omega)+\sqrt{\gamma_p}\hat{F}_{\gamma_p}^x(\omega)+\sqrt{\gamma_d}\hat{F}_{\gamma_d}^x(\omega)\big],
    \end{split}
\end{align}
% \begin{align}\label{eqn:NonIdealYout}
%     \begin{split}
%         \hat{Y}_{\text{out}}(\omega)&=\frac{-2g\sqrt{\kappa}(iJ)\Delta(\omega)+\big(i\omega+m_+\big)\big(i\omega+m_-)\delta\hat{Y}_{\text{in}}(\omega)+\big(\kappa/2\big)\sqrt{NC_{\text{eff}}\Gamma}\big[\sqrt{\gamma}\hat{Y}_{\gamma}(\omega)+\sqrt{\gamma_d}\hat{Y}_{\gamma_d}(\omega)\big]}{\big(i\omega-l_+\big)\big(i\omega-l_-\big)}\\[3pt]
%         &\hspace{2cm}+\frac{\big(\kappa/2\big)\sqrt{NC_{\text{eff}}\Gamma}}{\big(i\omega-l_+\big)\big(i\omega-l_-\big)}\big[\sqrt{\gamma}\hat{Y}_{\gamma}(\omega)+\sqrt{\gamma_d}\hat{Y}_{\gamma_d}(\omega)\big],
%     \end{split}
% \end{align}
where $l_{\pm}=\frac{\kappa+\Gamma}{4}\pm\sqrt{\big(\frac{\kappa-\Gamma}{4}\big)^2+2g^2Nz}$ and $m_{\pm}=\frac{\kappa-\Gamma}{4}\pm\sqrt{\big(\frac{\kappa+\Gamma}{4}\big)^2+2g^2Nz}$.
% \begin{align}
%     \begin{split}
%         \hat{Y}_{\text{out}}(\omega)&=-\frac{2g\sqrt{\kappa}(-iS)}{\big(i\omega-\kappa/2\big)\big(2i\omega-\Gamma+2 NC_{\text{eff}}\Gamma z\big)}\Delta(\omega)+\frac{\big(i\omega+\kappa/2\big)\big(2i\omega-\Gamma-2 NC_{\text{eff}}\Gamma z\big)}{\big(i\omega-\kappa/2\big)\big(2i\omega-\Gamma+2 NC_{\text{eff}}\Gamma z\big)}\delta\hat{Y}_{\text{in}}(\omega)\\[3pt]
%         &\hspace{2cm}-\frac{\kappa \sqrt{NC_{\text{eff}}\Gamma}}{\big(i\omega-\kappa/2\big)\big(2i\omega-\Gamma+2 NC_{\text{eff}}\Gamma z\big)}\big[\sqrt{\gamma}\hat{Y}_{\gamma}(\omega)+\sqrt{\gamma_d}\hat{Y}_{\gamma_d}(\omega)\big]
%     \end{split}
% \end{align}
 The smallest decay constant, $l_-$, is modified due to the presence of the extra dissipation sources and is always finite. In the limit $g\sqrt{N},\Gamma\ll \kappa$, these constants reduce to $l_+\approx m_+\approx \kappa/2$, $l_-\approx \Gamma(1-2NC_{\text{eff}}z)/2$ and $m_-\approx -\Gamma(1+2NC_{\text{eff}}z)/2$.
 
\begin{figure}
    \centering
    \includegraphics[width=\textwidth]{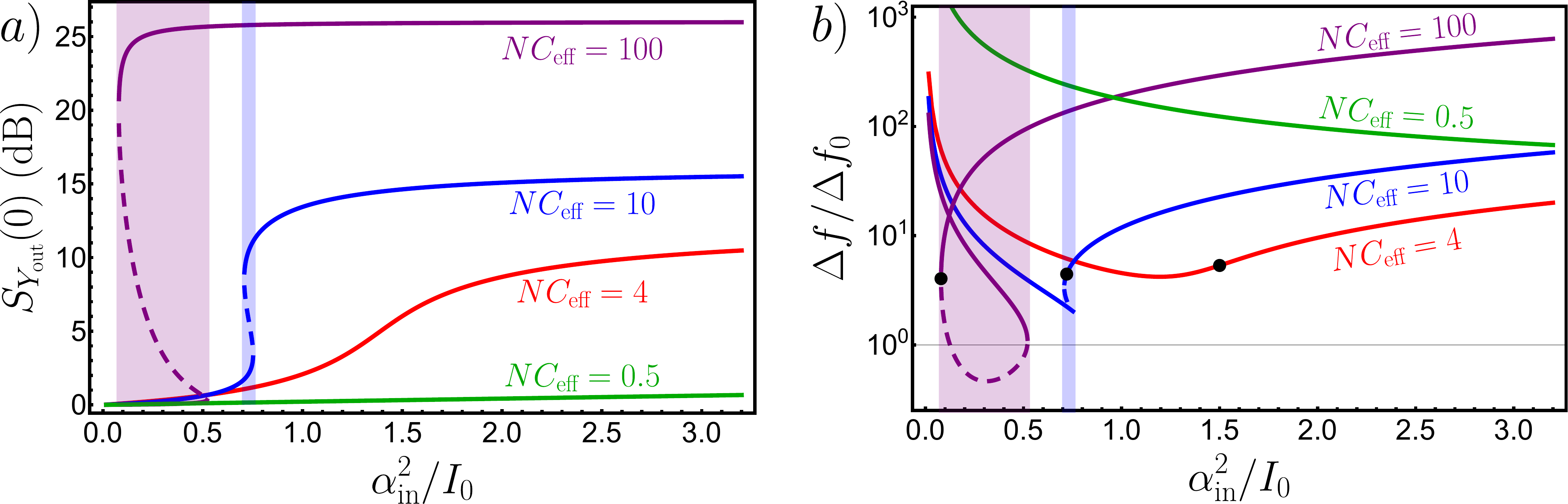}
    \caption{(a) Power spectral density of output quadrature $\hat{Y}_{\text{out}}$ as a function of input light intensity $\alpha_{\text{in}}^2$ for $NC_{\text{eff}}=0.5,4,10,100$ (green, red, blue and purple, respectively). Solid lines are stable solutions and dashed lines are unstable. Bistable regions are shaded. (b) Effective linewidth, measured in units of $\Delta\hspace{-0.05cm}f_0=(4\pi)^{-1}(C\gamma/4)[\Gamma/(\gamma_p+\gamma)]$. The "dark point" for each $NC_{\text{eff}}$ is marked by the black circle (nonexistent for the green curve).}
    \label{fig:NonIdealNoise}
\end{figure}

 The noise spectral density of $\hat{Y}_{\text{out}}$ at $\omega=0$ can be calculated from Eq.~(\ref{eqn:NonIdealYout}):
 \begin{align}\begin{split}\label{eqn:NonIdealEstimatorVariance}
     S_{Y_{\text{out}}}(0)=\frac{(1+2NC_{\text{eff}}z)^2}{(1-2N C_{\text{eff}}z)^2}+\frac{4NC_{\text{eff}}}{(1-2NC_{\text{eff}}z)^2}
 \end{split}\end{align}
 and is shown in Fig.~\ref{fig:NonIdealNoise}(a) as a function of $\alpha_{\text{in}}^2/I_0$. The absolute minimum noise improves with increasing $NC_{\text{eff}}$, but falls within bistable regions. Outside those regions, reducing $NC_{\text{eff}}$ brings the noise closer to shot noise levels.  
\subsection{Stabilization via feedback}
% In the presence of a constant frequency shift $\Delta(\omega)=2\pi \delta(\omega) \Delta_0$, a good estimator of $\Delta_t$ is $\hat{Y}_{\text{est}}=T^{-1}\int_0^T \hat{Y}_{\text{out}}(t)\, dt$. The average value of this estimator is modified with respect to the ideal case
% \begin{equation}
%     \langle\hat{Y}_{\text{est}}\rangle=\frac{4\alpha_{\text{in}}}{\Gamma}\Bigg[\frac{2NC_{\text{eff}}z}{(1-2NC_{\text{eff}}z)^2}\Bigg]\Delta_0.
% \end{equation}
% The variance is also modified due to the presence of the extra sources of noise. It is now given by
% \begin{equation}\label{eqn:NonIdealEstimatorVariance}
%     \langle\delta\hat{Y}_{\text{est}}^2\rangle=\frac{1}{4T}\Bigg[\frac{(1+2NC_{\text{eff}}z)^2}{(1-2N C_{\text{eff}}z)^2}+\frac{4NC_{\text{eff}}}{(1-2NC_{\text{eff}}z)^2}\Bigg],
% \end{equation}
% and thus the sensitivity is
% \begin{equation}\label{eqn:NonIdealLinewidth}
%     \langle\delta\Delta_0^2\rangle=\frac{C_{\text{eff}}\Gamma^2}{8\gamma T}\Bigg[\frac{(1+2NC_{\text{eff}}z)^2+4NC_{\text{eff}}}{(N C_{\text{eff}}+2NC_{\text{eff}}z)(-2NC_{\text{eff}}z)}\Bigg],
% \end{equation}
% where we have used Eq.~(\ref{eqn:NonIdealZ}) to get rid of $\alpha_{\text{in}}$ in favor of $z$. Note that, for fixed cavity and atomic parameters, this is still an implicit function of the input field intensity since $z$ depends on $\alpha_{\text{in}}$ and hence we are free to vary it to optimize this sensitivity (which is equivalent to optimizing over $z$). 
Applying the formulas of section~(\ref{sec:feed}) leads to the following effective linewidth:
\begin{equation}\label{eqn:NonIdealLinewidth}
    2\pi\Delta\hspace{-0.05cm}f=\Bigg(\frac{C\gamma}{4}\Bigg)\Bigg(\frac{\gamma+\gamma_p+\gamma_d}{\gamma+\gamma_p}\Bigg)\Bigg[\frac{(1+2NC_{\text{eff}}z)^2+4NC_{\text{eff}}}{2(N C_{\text{eff}}+2NC_{\text{eff}}z)(-2NC_{\text{eff}}z)}\Bigg],
\end{equation}
which is parameterized by $z$ and hence is an implicit function of the input field intensity $\alpha_{\text{in}}^2$ (since $z$ depends on $\alpha_{\text{in}}$). We have expressed Eq.~(\ref{eqn:NonIdealLinewidth}) as a product of three different factors to make the discussion clearer. The first factor is the scaling present both in the ideal section and in the superradiant laser. The second factor $\Gamma/(\gamma+\gamma_p)$ strongly penalizes the effective linewidth when $\gamma_d\gg \gamma+\gamma_p$, which is a consequence of a reduction in the size of the radiating dipole moment $J\propto \sqrt{(\gamma+\gamma_p)/\Gamma}$ [see Eq.~(\ref{eqn:NonIdealSS})]. Under these conditions, an adjustable depumping rate provides an advantage when the natural $\gamma$ is small, as is the case in narrow linewidth transitions. If $\gamma_p\gg \gamma_d$, then $(\gamma+\gamma_p)/\Gamma \approx 1$ and there is no penalization. The third factor includes all the effects of operating at different $\alpha_{\text{in}}$ and is difficult to estimate a priori. It can be of order $1$ or much bigger than $1$ depending on $\alpha_{\text{in}}$, as shown in the next subsection.

In Fig.~\ref{fig:NonIdealNoise}(b) we show $\Delta\hspace{-0.05cm}f$ as a function of input light intensity for different values of $NC_{\text{eff}}$, normalized to $\Delta\hspace{-0.05cm}f_0=(4\pi)^{-1}(C\gamma/4)[\Gamma/(\gamma+\gamma_p)]$. When $NC_{\text{eff}}>8$, the $\alpha_{\text{in}}$ that minimizes $\Delta f$ falls within regimes of bistability. Outside of these regions, we find that it is preferable to work at more moderate values of $NC_{\text{eff}}$ for fixed $\alpha_{\text{in}}^2/I_0$. Reducing $NC_{\text{eff}}$ too much is not convenient either because the size of the signal acquired through the output light is also reduced, negatively impacting $\Delta\hspace{-0.05cm}f$ [see Fig.~\ref{fig:NonIdealNoise}(b) for $NC_{\text{eff}}=0.5$].
\subsection{Operating regimes}
We describe here two possible operating regimes corresponding to different values of $\alpha_{\text{in}}$, discuss the achievable effective linewidths $\Delta f$, and calculate the corner frequencies $\omega_S$ and $\omega_R$.
%$z\approx-\frac{I_0}{16 (NC_{\text{eff}})^2\alpha_{\text{in}}^2}\to 0$
%[see Eq.~(\ref{eqn:NonIdealEstimatorVariance}) at $z=0$]
\subsubsection{Strong input field}
Here we take $\alpha_{\text{in}}^2\gg I_0=N^2C_{\text{eff}}(\gamma_p+\gamma)/16$, a situation previously analyzed in Ref.~\cite{Martin2011}, and which leads to $z\sim 1/\alpha_{\text{in}}^2\to 0$, $\alpha_{\text{out}}\approx -\alpha_{\text{in}}$, and $J$ (dipole moment) $\sim 1/\alpha_{\text{in}}$. Since $J$ shrinks with larger $\alpha_{\text{in}}$, the size of the signal necessary to estimate frequency fluctuations is reduced, while the noise in the output light does not change significantly [see Eq.~(\ref{eqn:NonIdealEstimatorVariance}) at $z=0$]. The effective linewidth is:
\begin{equation}
    2\pi\Delta \hspace{-0.005cm}f=\frac{C\gamma}{4}\Bigg(\frac{\Gamma}{\gamma_p+\gamma}\Bigg)\Bigg(\frac{\alpha_{\text{in}}^2}{4I_0}\Bigg)\big(1+4NC_{\text{eff}}\big)
\end{equation}
In this case, apart from $\Gamma/(\gamma+\gamma_p)$, there is an extra factor proportional to the input intensity that is much larger than one by assumption. This is the linewidth found in Ref.~\cite{Martin2011}, except for the factor $(1+4NC_{\text{eff}})$ which includes the additional noise due to the single particle decoherence processes. Even when $NC_{\text{eff}}>8$, the strong input field region is free from bistability, which is advantageous if manipulating $NC_{\text{eff}}$ is not possible or easy. On the other hand, the achievable effective linewidth is then also penalized from the extra $(1+4NC_{\text{eff}})$ factor.

In contrast to the ideal case, $\omega_S$ and $\omega_R$ are now different. In the limit $g\sqrt{N},\Gamma\ll \kappa$ they are $\omega_S=\Gamma\sqrt{NC_{\text{eff}}+1/4}$ and $\omega_R=\Gamma/2$. It is desirable to have a large $\omega_S$, either by increasing $NC_{\text{eff}}$, at the expense of a larger linewidth $\Delta\hspace{-0.05cm}f$, or by modifying the depumping rate $\gamma_p$. An important final piece of information is an estimate of the size of detunings $\Delta_t$ under which the linear approximation is valid. Solving the mean field equations in the presence of a static detuning $\Delta_0$ leads to a dispersive curve similar to Fig.~\ref{fig:IdealSystem}(a), but where the maximal value is obtained at $\Delta_0\approx NC_{\text{eff}}\Gamma\alpha_{\text{in}}/\sqrt{8I_0}=NC\gamma\alpha_{\text{in}}/\sqrt{8I_0}$ instead of $\Delta_0\approx NC\gamma$ (ideal case). Linearity is thus valid for a larger range of $\Delta_t$, as compared to the ideal case, at the expense of a smaller slope and hence lower sensitivity.

\subsubsection{Dark point}
This corresponds to no light coming outside the cavity, namely $\alpha_{\text{out}}=0$.
%Here, the mean field observables take the values
% \begin{equation}
%     \alpha_{\text{in}}=\sqrt{\gamma}\sqrt{\frac{NC_{\text{eff}}-1}{2C_{\text{eff}}}} \hspace{0.75cm}z=-\frac{1}{2NC_{\text{eff}}}\hspace{0.75cm} S=-iN\sqrt{\frac{\gamma}{\Gamma}}\sqrt{\frac{NC_{\text{eff}}-1}{(NC_{\text{eff}})^2}}\hspace{0.75cm}\alpha=\frac{\alpha_{\text{in}}}{\sqrt{\kappa}}.
% \end{equation}
At this operating point, the effective linewidth is:
\begin{equation}\label{eqn:darkpointlw}
    2\pi\Delta \hspace{-0.05cm}f=\frac{C\gamma}{2}\Bigg(\frac{\Gamma}{\gamma+\gamma_p}\Bigg) \Bigg(\frac{NC_{\text{eff}}}{NC_{\text{eff}}-1}\Bigg).
\end{equation}
and only exists when $NC_{\text{eff}}>1$. In this case, apart from $\Gamma/\gamma$, there is a third factor which is of order one when $NC_{\text{eff}}\gtrsim 2$. Hence, the achievable linewidth can be of the same size as that of the superradiant laser ($\sim C\gamma$)~\cite{Meiser2009}, provided $\gamma+\gamma_p$ and $\Gamma$ are comparable. Consider again the $^1S_0\to\,^3P_0$ transition in $^{87}\text{Sr}$ in the cavity of Ref.~\cite{norcia2018cavity}. Using $g=2\pi\times 4$ Hz, $\kappa=2\pi\times 160$ kHz, $\gamma_d=2\pi\times 3$ Hz, an artificially augmented spontaneous emission rate $\gamma_p+\gamma=\gamma_d$ and $N=10^5$ atoms leads to $\Delta\nu=0.5$ mHz at $NC_{\text{eff}}\approx 6.6$.

Within the approximations $g\sqrt{N},\Gamma\ll\kappa$, the corner frequencies are now $\omega_S=\sqrt{NC_{\text{eff}}\Gamma^2}=\sqrt{NC\gamma \Gamma}$ and $\omega_R=\Gamma$. Increasing $\omega_S$ while avoiding bistability ($NC_{\text{eff}}<8$) requires increasing the depumping rate $\gamma_p$. Furthermore $\gamma_p$ also enhances $\omega_R$, which could be as small as $\gamma$ when $\gamma_d=\gamma_p=0$, and this relaxes technical constraints on the feedback loop gain. Note also that if we apply a depumping strong enough that $\Gamma\sim NC\gamma/2$, then $\omega_S \sim NC\gamma$, similar to the result for the superradiant laser at optimal pumping rate. Finally, the response ($\hat{Y}_{\text{out}}$) to large static detunings $\Delta_0$ is still dispersive, as in Fig.~\ref{fig:IdealSystem}(a), with the maximal value attained at $\Delta_0\approx NC_{\text{eff}}\Gamma=NC\gamma$ (similar to ideal case).

This ``dark point" is representative of the properties of the system close to the minimum $\Delta\hspace{-0.05cm} f$ [see Fig.~\ref{fig:NonIdealNoise}(b)], and could be easily identified by looking at the reflected intensity. Numerically, we find that Eq.~(\ref{eqn:darkpointlw}) is at most within a factor of $2$ of the optimal $\Delta\hspace{-0.05cm} f$ when $2<NC_{\text{eff}}<8$. 

% \subsubsection{Optimal point}
% The expression in Eq.~(\ref{eqn:NonIdealLinewidth}) can be optimized over $z$, leading to
% \begin{align}\begin{split}
%     % z&=-\frac{1}{2}\Bigg(\frac{\sqrt{1+4NC_{\text{eff}}}}{1+NC_{\text{eff}}+\sqrt{1+4NC_{\text{eff}}}}\Bigg)\\
%     2\pi\Delta \hspace{-0.05cm}f&=(\delta\Delta_0^2) T=\frac{C\gamma}{2}\Bigg(\frac{\Gamma}{\gamma+\gamma_p}\Bigg)\frac{\Big(1+\sqrt{1+4NC_{\text{eff}}}\Big)\Big(2NC_{\text{eff}}+1+\sqrt{1+4NC_{\text{eff}}}\Big)}{(2NC_{\text{eff}})^2}
% \end{split}\end{align}
% The linewidth shows again the ideal scaling $g^2/\kappa$, the penalizing factor $\Gamma/\gamma$, and a factor that depends on $NC_{\text{eff}}$. This last factor decreases monotonically from $2$ to $0.6$ in the region $2<NC_{\text{eff}}<8$, so contributes an order one correction to the linewidth in the parameter regimes without bistability. Using the same parameters as before leads to $\Delta\nu=0.3$ mHz. When $NC_{\text{eff}}\gg 1$, the extra factor approaches $0$ as $1/\sqrt{NC_{\text{eff}}}$, but falls within the bistable region and is part of the unstable solution branch, so inaccessible.
\section{Conclusions and outlook}
We have calculated the effective linewidth of a laser stabilized to an ensemble of atoms inside a QED cavity across various values of input light intensity and atom-light interaction strength, measured in terms of the cooperativity parameter $NC_{\text{eff}}$. We find that working with moderate values of $NC_{\text{eff}}\in [2,8]$ is preferable and can lead to effective linewidths comparable to those of the superradiant laser, provided the total dephasing and effective spontaneous emission rates are of the same size. These predictions open a path for a complementary implementation of ultranarrow linewidth light which does not require a strong incoherent pump but instead coherent drive and feedback and thus a more versatile and amenable avenue for the experimental generation of continuous coherent light with long lived dipoles. Furthermore, concerns about the finite lifetime of atoms in the cavity can be addressed by continuous atom loading schemes~\cite{Cline2022}.

The achievable linewidths using this configuration in state-of-the-art cavities with alkaline-earth atoms are in the mHz range. Further improvements would require reducing the ratio $g^2/\kappa$ while keeping $NC_{\text{eff}}$ fixed, which could be realized using a larger number of atoms. 

%When $NC_{\text{eff}}\ll 1$ the noise in the output light is at the level of shot noise, but the size of the signal necessary for feedback is greatly reduced. When $NC_{\text{eff}}\gg 1$, outside regions of bistability the signal remains small and the noise in the output light also increases.

\section{Acknowledgements}
We thank E. Y. Song and A. Ludlow for a careful reading and comments on the manuscript, and J. R. K. Cline and D. J. Young for helpful discussions. This work is supported by the AFOSR Grant No. FA9550-18-1-0319, by the DARPA (funded via ARO) Grant No. W911NF-16-1-0576, the ARO single investigator Grant No. W911NF-19-1-0210, the NSF PHY1820885, NSF JILA-PFC PHY-1734006 and NSF QLCI-2016244 grants, by the DOE Quantum Systems Accelerator (QSA) grant and by NIST.
\begin{center}
\textbf{Table 1}. List of parameters I\vspace{1em}
\begin{tabular}{ | >{\centering\arraybackslash}m{1em}  m{14em} |>{\centering\arraybackslash}m{9.5em} m{15.5em}|} 
  \hline
  $g$ & Single photon Rabi frequency & $\omega_a$, $\omega_c$, $\omega_d$ & Atomic, cavity, drive frequencies\\
  
  $\kappa$ & Cavity power decay rate & $\Delta_t=\omega_d-\omega_a$ & Drive-atom detuning\\
  
  $\gamma$&Spontaneous emission rate & $\Gamma=\gamma+\gamma_p+\gamma_d$& Total dephasing rate\\
  
  $\gamma_p$& Depumping rate &$C=4g^2/(\kappa\gamma)$& Cavity cooperativity\\ 
  
  $\gamma_d$& Dephasing rate &$C_{\text{eff}}=C\gamma/\Gamma$ & Effective cooperativity\\ 
  
  $N$ & Atom number& $\alpha_{\text{in}}^c=gN/(2\sqrt{\kappa})$ & Ideal critical input field\\ 
  
  $\Delta f$ & Effective linewidth &$I_0=(\alpha_{\text{in}}^c)^2(\gamma+\gamma_p)/\Gamma$& Input field scale\\
  \hline
\end{tabular}
\vspace{1em}
\end{center}

\begin{center}
\textbf{Table 2.} List of parameters and variables II ($O$ stands for any variable)\vspace{1em}
\begin{tabular}{ | >{\centering\arraybackslash}m{10em}  >{\centering\arraybackslash}m{10em} |>{\centering\arraybackslash}m{10em} >{\centering\arraybackslash}m{13em}|} 
  \hline
  $\hat{a}=\hat{X}+i\hat{Y}$ & Intracavity field & $\alpha$ & Mean field intracavity field\\
  
  $\hat{S}_z$, $\hat{S}^-=\hat{S}_x-i\hat{S}_y$ & Atomic variables &$Z=Nz, J$ & Mean field atomic variables\\
  
  $\hat{A}_{\text{in}}=\hat{X}_{\text{in}}+i\hat{Y}_{\text{in}}$& Input field& $\alpha_{\text{in}}$ & Mean field input field\\
 
  $\hat{A}_{\text{out}}=\hat{X}_{\text{out}}+i\hat{Y}_{\text{out}}$& Output field& $\alpha_{\text{out}}$ & Mean field output field\\
  \hline
  \vspace{0.25em}
  $\hat{F}_{\gamma}^z$, $\hat{F}_{\gamma}^-=\hat{F}_{\gamma}^x-i\hat{F}_{\gamma}^y$& Noise due to $\gamma$&  $\hat{F}_{\gamma_{d}}^-=\hat{F}_{\gamma_{d}}^x-i\hat{F}_{\gamma_{d}}^y$& Noise due to $\gamma_d$ \\

  $\hat{F}_{\gamma_p}^z$, $\hat{F}_{\gamma_p}^-=\hat{F}_{\gamma_p}^x-i\hat{F}_{\gamma_p}^y$& Noise due to $\gamma_p$ &$\hat{S}_O(\omega)$& Noise spectral density of $O$\\
 
  \hline
  \vspace{0.25em}
  $R(\omega)$& \hspace{-3em}Response function Eq.~(\ref{eqn:Response})&$O(\omega)=\int e^{i\omega t} O(t)\,dt$ & Fourier transform of $O$ \\
  $\omega_S$ & \hspace{-3em}Corner frequency of $S_{\Delta}(\omega)$&$\omega_R$ & Corner frequency of $R(\omega)$\\
  \hline
\end{tabular}
\end{center}

\bibliography{library}

\end{document}